ORIGINAL ARTICLE

# Controlling light-with-light without nonlinearity

Jianfa Zhang, Kevin F MacDonald and Nikolay I Zheludev

According to the fundamental Huygens superposition principle, light beams traveling in a linear medium will pass though one another without mutual disturbance. Indeed, the field of photonics is based on the premise that controlling light signals with light requires intense laser fields to facilitate beam interactions in nonlinear media, where the superposition principle can be broken. Here we challenge this wisdom and demonstrate that two coherent beams of light of arbitrarily low intensity can interact on a metamaterial layer of nanoscale thickness in such a way that one beam modulates the intensity of the other. We show that the interference of beams can eliminate the plasmonic Joule losses of light energy in the metamaterial or, in contrast, can lead to almost total absorption of light. Applications of this phenomenon may lie in ultrafast all-optical pulse-recovery devices, coherence filters and terahertz-bandwidth light-by-light modulators.

## INTRODUCTION

In 1678, Christiaan Huygens stipulated that '…light beams traveling in different and even opposite directions pass though one another without mutual disturbance'[1] and in the framework of classical electrodynamics, this superposition principle remains unchallenged for electromagnetic waves interacting in vacuum or inside an extended medium.[2] Since the invention of the laser, colossal effort has been focused on the study and development of intense laser sources and nonlinear media for controlling light with light, from the initial search for optical bistability[3] to recent quests for all-optical data networking and silicon photonic circuits. However, interactions of light with nanoscale objects provide some leeway for violation of the linear superposition principle. This is possible through the use of coherent interactions, which have been successfully engaged in applications ranging from phased array antennas to the manipulation of light distributions and quantum states of matter.[4–11]

Consider a thin light-absorbing film of sub-wavelength thickness: the interference of two counter-propagating incident beams A and B on such a film is described by two limiting cases illustrated in Figure 1: in the first, a standing wave is formed with a zero-field node at the position of the absorbing film. As the film is much thinner than the wavelength of the light, its interaction with the electromagnetic field at this minimum is negligible and the absorber will appear to be transparent for both incident waves. On the other hand, if the film is at a standing wave field maximum, an antinode, the interaction is strong and absorption becomes very efficient. Altering the phase or intensity of one beam will disturb the interference pattern and change the absorption (and thereby transmission) of the other. For instance, if the film is located at a node of the standing wave, blocking beam B will lead to an immediate increase in loss for beam A and therefore a decrease in its transmitted intensity. Alternatively, if the film is located at an antinode of the standing wave, blocking beam B will result in a

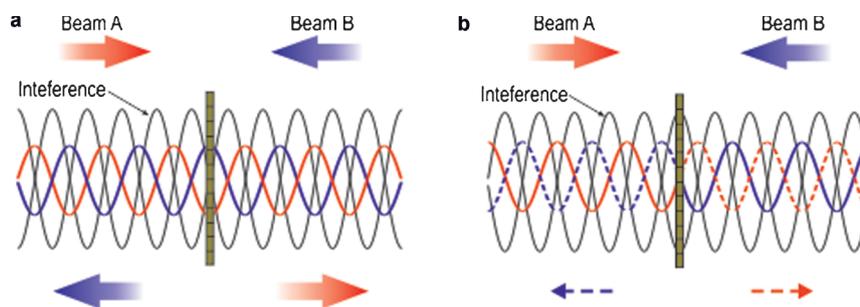

**Figure 1** Interaction of light with light on a nanoscale absorber. Two coherent counter-propagating beams A and B are incident on an absorber of sub-wavelength thickness, for instance, a lossy plasmonic metamaterial film. Two limiting regimes of interaction exist wherein the beams at the film interfere either (**a**) destructively or (**b**) constructively to effect total transmission or total absorption, respectively.

Optoelectronics Research Centre & Centre for Photonic Metamaterials, University of Southampton, Southampton, Hampshire SO17 1BJ, UK
Correspondence: Professor NI Zheludev, Optoelectronics Research Centre, University of Southampton, Highfield, Southampton, Hampshire SO17 1BJ, UK
E-mail: niz@orc.soton.ac.uk

decrease of losses for beam A and an increase in its transmitted intensity. In short, manipulating either the phase or intensity of beam B modulates the transmitted intensity of beam A.

To optimize the modulation efficiency, the film should absorb half of the energy of a single beam passing through it. Under such circumstances, 100% light-by-light modulation can be achieved when signal A is modulated by manipulating the phase of beam B and 50% modulation can be achieved if control is encoded in the intensity of beam B. Moreover, one will observe that when the intensities of the two beams are equal and the film is located at an antinode, all light entering the metamaterial will be absorbed, while at a node, light transmitted by the film will experience no Joule losses.

Here, it should be noted that for fundamental reasons, an infinitely thin film can absorb not more than half of the energy of the incident beam.[12,13] At the same time, a level of absorption of 50% is difficult to achieve in thin unstructured metal films: across most of the optical spectrum, incident energy will either be reflected or transmitted by such a film. Recently reported much higher absorption levels have only been achieved in layered structures of finite thickness[14–18] that are unsuitable for implementation of the scheme presented in Figure 1. However, in the optical part of the spectrum, a very thin nanostructured metal film can deliver strong resonant absorption approaching the 50% target at a designated wavelength. Such metal films, periodically structured on the sub-wavelength scale, are known as planar plasmonic metamaterials.

## MATERIALS AND METHODS

The experimental arrangement presented in Figure 2 was employed to demonstrate light-by-light modulation and total absorption/transparency for a plasmonic metamaterial. A linearly polarized beam of light from a HeNe laser (wavelength $\lambda=632.8$ nm) is divided by a pellicle beam-splitter BS1 into two beams A and B, denoted as 'signal' and 'control' beams, respectively, which are adjusted to equal intensity by an attenuator in path B. The beams are focused at normal incidence onto the plasmonic metamaterial (PMM) from opposing directions by parabolic mirrors. The phase of control beam B is manipulated via a piezoelectrically actuated optical delay line while a mechanical chopper provides for modulation of its intensity. The intensities of the beams transmitted by the metamaterial are monitored by a single photodetector, which may register the combined intensity of both beams (the difference in path length from metamaterial to detector for the two beams being much longer than the coherence length of the laser radiation so there is no optical interfere at the detector) or that of either single beam (the other being shuttered accordingly).

As the key element for light-by-light modulation, we employ a metamaterial with a thickness of $\lambda/13$—a two-dimensional array of asymmetric split-ring plasmonic resonators milled through a 50-nm gold film. This nanostructure supports a Fano-type plasmonic mode[19,20] that leads to strong resonant absorption. The pitch of metamaterial array, with a unit cell size of 250 nm×250 nm, is smaller than the wavelength and as such, it does not diffract light. Details of metamaterial structure and fabrication and the dispersion of its optical properties may be found in the Supplementary Information.

## RESULTS AND DISCUSSION

Figure 3 illustrates the modulation of signal intensity via manipulation of control beam phase ϕ (Figure 3a and 3b) or intensity (Figure 3c). The phase of the control beam is changed using the delay line in arm B. Continuously changing the phase has the effect of translating the metamaterial film between nodes (ϕ=π, 3π, …) and antinodes (ϕ=0, 2π, …) of the standing wave, bringing about a modulation of the detected signal (channel A) intensity between levels at 115% and 10% of the incident level (Figure 3a).

For an ideal, free-standing, zero-thickness 50% absorber, one would see the signal beam modulated between 0% and 100% of the incident intensity level. The somewhat different limits between which experimental modulation is observed are explained by a number of factors: First, the sample's absorption level at the laser wavelength is not exactly 50%. Indeed, due to the presence of a substrate and to fabrication-related asymmetry/imperfection of the slots milled into the gold film, it shows differing levels of absorption (34% and 57%) for the two opposing propagation directions; Second, although the metamaterial is very thin it does have a finite thickness of $\lambda/13$; and finally, the laser source is not perfectly coherent—its emission includes an incoherent luminescence component (detailed optical characterization and computational modeling of the experimental sample are presented in the Supplementary Information).

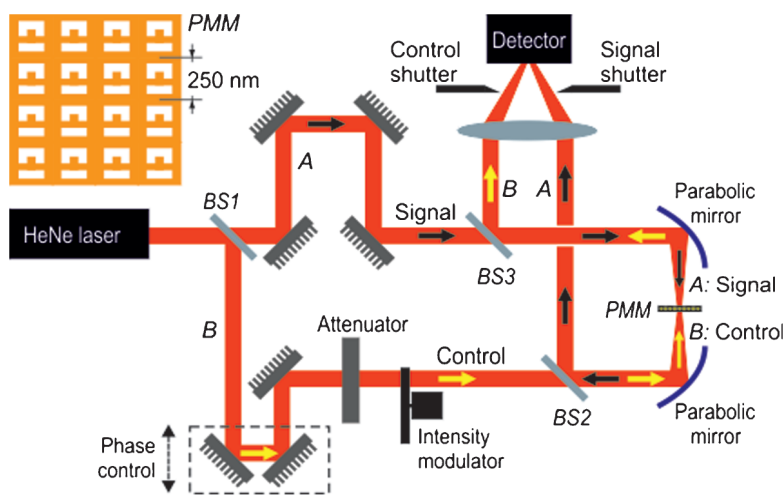

**Figure 2** Experimental arrangement for demonstration of optically-controlled transparency/absorption in a plasmonic metamaterial. Black and yellow arrows respectively indicate the paths of 'signal' and 'control' beams A and B from beam-splitter BS1, where they are generated from a single source beam, to the plasmonic metamaterial PMM and from there to the photodetector. Changes in the phase or intensity of control beam B can be used to switch between regimes of total transmission and total absorption at the metamaterial, and so to modulate the intensity of signal beam A. The inset shows a schematic of the metamaterial—a gold film perforated with an array of asymmetric split-rings (see Supplementary Information for further detail of sample structure).

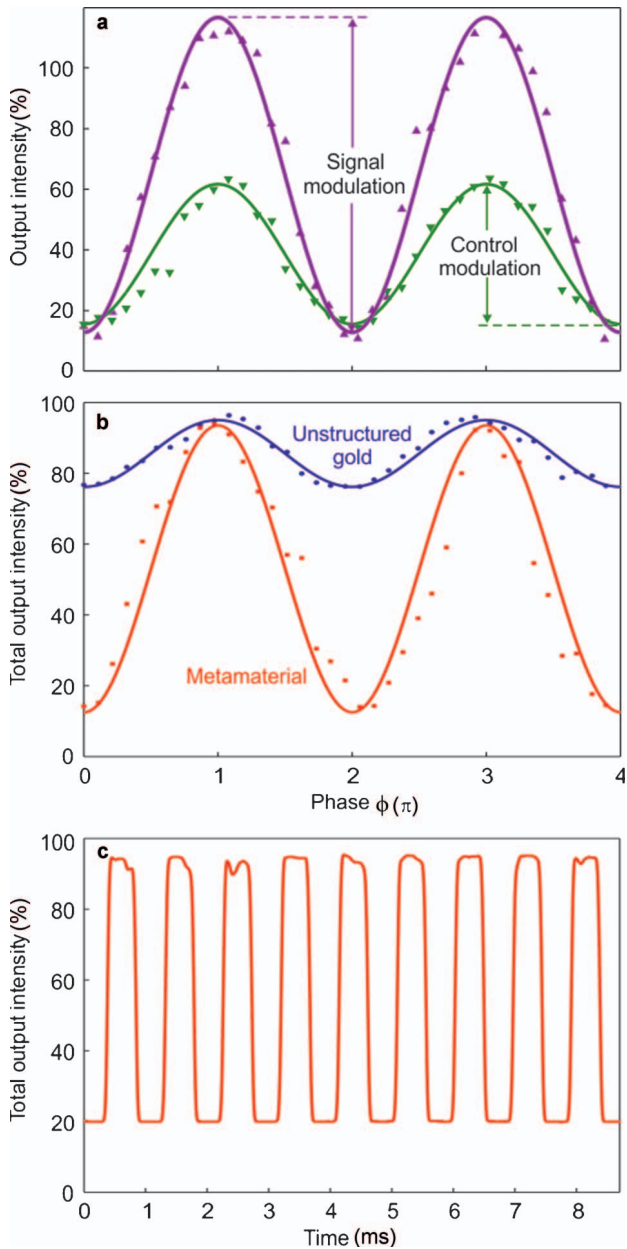

**Figure 3** Controlling light-with-light in a plasmonic metamaterial. (**a**) Output intensity in signal and control channels, relative to their respective input intensities, as functions of the mutual phase of the input beams at the metamaterial film. (**b**) Combined output intensity of the signal and control channels, relative to combined input intensity, as a function of the mutual phase of the incident beams for the plasmonic metamaterial absorber and for an unstructured 50-nm gold film. (**c**) Time domain trace of output (combined intensity) modulation effected via intensity modulation of the input control beam (mechanically chopped at 1.07 kHz).

Figure 3b compares phase modulation of total output intensity for the metamaterial array and an unstructured area of the same gold film. The modulation amplitude for the unstructured metal is severely limited because it is highly reflective and therefore presents single-beam absorption far below the optimal 50% level. Through metamaterial patterning, one can manipulate the balance among absorption, transmission and reflection in sub-wavelength plasmonic films to achieve desired levels at any visible/infrared wavelength.

Figure 3c shows modulation of total output intensity resulting from modulation of control intensity in the time domain. When the control beam is blocked, only the signal wave is present at the metamaterial and the standing wave regime of light–metamaterial interaction is replaced by the traveling wave regime: In this example, the metamaterial is initially located at a node of the standing wave where absorption is minimal (combined output intensity=95% of input); interruption of the control beam 'switches on' signal beam absorption and output drops to 20%. This proof-of-principle demonstration employs a mechanical chopper running at only 1.07 kHz. However, we argue that the cross-beam modulation bandwidth will be limited only by the width of the resonant absorption peak, and as such may be in the terahertz range (see below).

To further illustrate the potential for application of coherent control over metamaterial absorption in real-world devices, we consider the performance of a free-standing (no substrate) 50-nm gold metamaterial film with an absorption line engineered for the telecommunications band centered at 1550.5 nm (Figure 4). The metamaterial, modeled on the basis of well-established data for the complex conductivity of gold,[21] exhibits single-beam absorption of 50.18%. As such, it will deliver phase-controlled total absorption of between 0.38% and 99.99% and total output intensity modulation between levels of 99.62% and 0.01% of total (combined) input intensity. The relatively broad nature of the metamaterial resonance provides for modulation between 1% and 90% of input intensity levels across the entire spectral range from 1530 to 1575 nm, giving a bandwidth of 5.6 THz.

We consider that the potential applications of the effect are manifold and of considerable technological importance. The high sensitivity of absorption to the mutual phase of beams may be harnessed for applications in sensors and the effect may find use in laser spectroscopy. However, the most striking applications may lie in the domain of signal processing (Figure 5), for example in:

- Photonic 'pulse restoration' or 'clock recovery' (Figure 5a). In optical data systems, signal pulses become distorted through dispersion and nonlinear interactions, slowing down data distri-bution and processing. A distorted pulse may be 'cleaned up' through interaction with a clock pulse at a nanoscale metamaterial absorber. Indeed, in the total transmission regime, spectral components of the distorted pulse that have the same intensity and amplitude as the clock pulse will be transmitted with negligible loss, while distorted components are strongly absorbed, thereby restoring the temporal and spectral profile of the signal.
- Coherence filtering (Figure 5b). Following the same principle as behind 'pulse restoration', i.e., that the absorption of the coherent part of a signal can be enhanced or eliminated, one may realize a filter with the extraordinary ability to increase or decrease the mutual coherence of two light beams.[22]
- Optical gating (Figure 5c). Coherent control of absorption provides functionality for analog and digital, all-optical (light-by-light) modulation/switching without any optically nonlinear medium, thereby delivering this functionality at extremely low power levels. The coherent control approach requires a coherent information carrier across the entire network, but promises extremely high, terahertz frequency modulation bandwidth, which is determined by the width of metamaterial plasmonic resonance. Using plasmonic metal nanostructures, the approach may be implemented across the entire visible and near-infrared spectral range, where resonances can be engineered by design and metallic Joule losses are substantial. It should be noted that while metamaterial optical gates have a truth table that can also be obtained using a conventional interferometer, they deliver this by manipulating the absorption of energy rather than transferring in between output ports. The planar metamaterial implementation may also present advantages for the assembly of compact, highly-integrated (cascaded) devices. Finally, in certain data processing applications, metamaterial absorbers of finite sub-wavelength thickness exhibiting absorption higher than 50% may offer useful solutions.

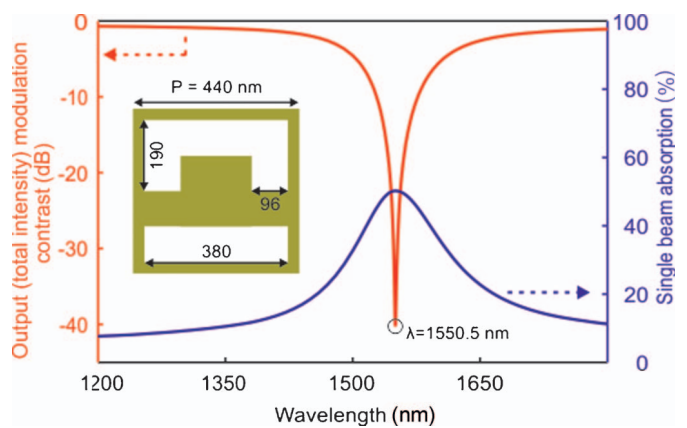

**Figure 4** Metamaterial modulator for telecommunications. Dispersion of the cross-intensity modulation contrast (red) and traveling wave absorption (blue) of a plasmonic metamaterial engineered to function as a coherently controlled absorber at 1550.5 nm (the metamaterial unit cell geometry is shown inset; a free-standing gold film thickness of 50 nm is assumed).

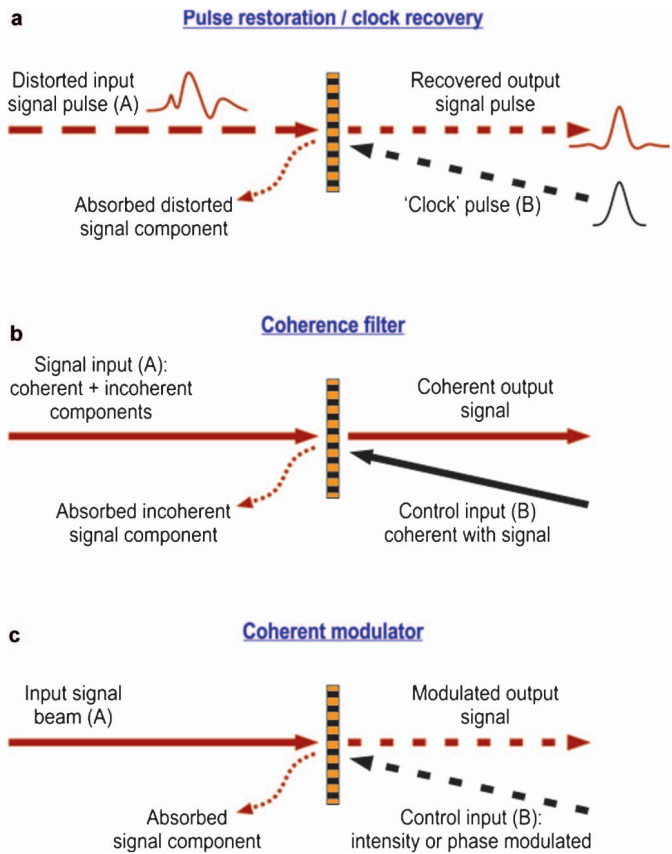

**Figure 5** Applications. (**a**) A pulse restoration (clock recovery) device to restore the form of distorted signal pulses according to that of a clock (control) pulse; (**b**) a coherence filter that improves the coherence of light beams by absorbing incoherent components; (**c**) a coherent light-by-light modulator wherein a digital or analogue intensity- or phase-modulated control input governs signal channel output.

## CONCLUSIONS

In summary, we have demonstrated for the first time that a plasmonic metamaterial—a single layer of nanostructured metal much thinner than the wavelength of light, can be used to modulate light with light.

Regimes of near-total absorption and near-total suppression of plasmonic losses have been experimentally observed. The phenomenon relies on the coherent interaction of light beams on the metamaterial and provides functionality that can be implemented freely across a broad visible to infrared range by varying the structural design. It may serve applications in sensors, variable attenuators, unique light coherence filters and terahertz-bandwidth pulse-recovery devices that can operate at extremely low power levels.

## ACKNOWLEDGMENTS

The authors thank Jun-Yu Ou and Mengxin Ren for assistance with nanofabrication and optical experiments respectively. This work was supported by the Engineering and Physical Sciences Research Council (grant EP/G060363/1), The Royal Society and the China Scholarship Council.

Supplementary Information for this article can be found on *Light: Science & Applications*' website (http://www.nature.com/lsa/).